\documentclass[%
 reprint,
 amsmath,amssymb,
 sd,
notitlepage
]{revtex4-1}
\usepackage{hyperref}
\usepackage{graphicx}
\usepackage{dcolumn}
\usepackage{bm}
\usepackage{enumitem}
\usepackage{bigints}
\usepackage[autostyle]{csquotes}
\usepackage[T1]{fontenc}
\usepackage[english]{babel}
\MakeOuterQuote{"}

\newcommand{\pr}{^\prime}
\newcommand{\dprime}{^{\prime\prime}}
\newcommand{\dd}[1]{{\mathrm{d}}#1\,}

\newcommand{\br}[1]{\langle #1 \vert}
\newcommand{\ke}[1]{\vert #1 \rangle}

\newcommand{\ev}[1]{\langle #1 \rangle}
\renewcommand{\vec}[1]{\boldsymbol{\mathbf{#1}}}

\usepackage{xcolor}

\begin{document}

\title{Probing Nonadiabatic Dynamics with Attosecond Pulse Trains and Soft X-ray Raman Spectroscopy}

\author{Lorenzo Restaino}

 \affiliation{Department of Physics, Stockholm University, Albanova University Centre, SE-106 91 Stockholm, Sweden}
 \author{Deependra Jadoun}%
\affiliation{Department of Physics, Stockholm University, Albanova University Centre, SE-106 91 Stockholm, Sweden}
\author{Markus Kowalewski}%
 \email{e-mail: markus.kowalewski@fysik.su.se}
\affiliation{Department of Physics, Stockholm University, Albanova University Centre, SE-106 91 Stockholm, Sweden}

\begin{abstract}
Linear off-resonant X-ray Raman techniques are capable of detecting the ultrafast electronic coherences generated when a photoexcited wave packet passes through a conical intersection.
A hybrid femtosecond or attosecond probe pulse is employed to excite the system and stimulate the emission of the signal photon, where both fields are components of a hybrid pulse scheme.
In this paper, we investigate how attosecond pulse trains, as provided by high-harmonic generation processes, perform as probe pulses in the framework of this spectroscopic technique, instead of single Gaussian pulses.
We explore different combination schemes for the probe pulse, as well as the impact of parameters of the pulse trains on the signals. Furthermore, we show how Raman selection rules and symmetry consideration affect the spectroscopic signal, and we discuss the importance of vibrational contributions to the overall signal.
We use two different model systems, representing molecules of different symmetry, and quantum dynamics simulations to study the difference in the spectra. The results suggest that such pulse trains are well suited to capture the key features associated with the electronic coherence.
\end{abstract}
\maketitle

\section{Introduction}\label{introduction}
\begin{figure}
    \includegraphics[width=\linewidth]{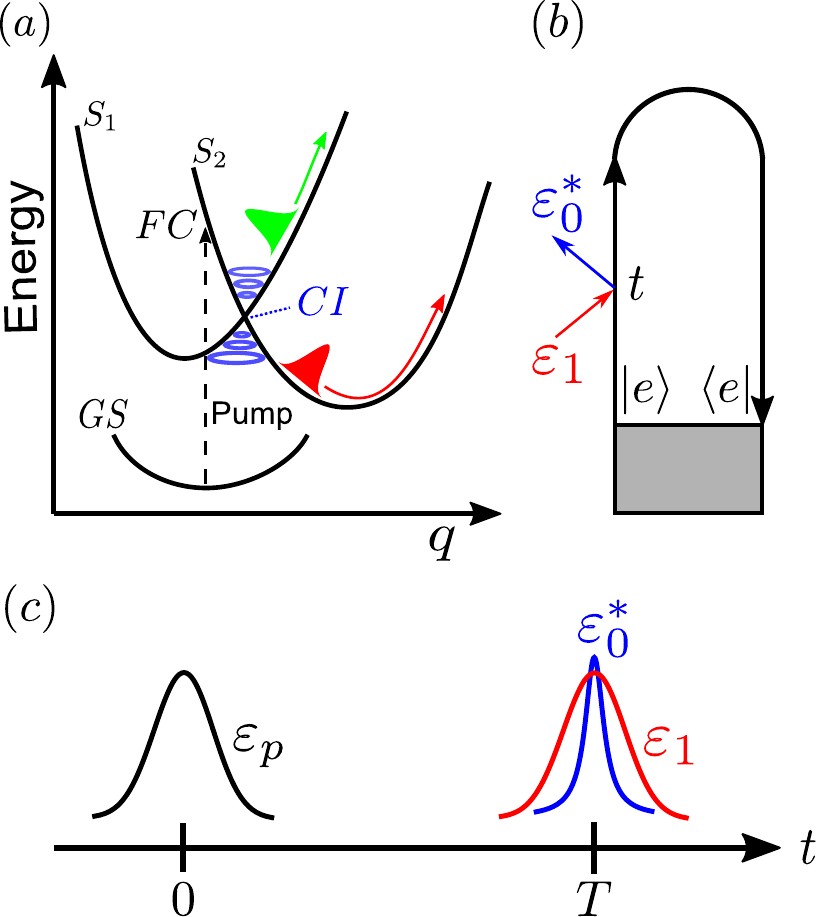}
    \caption{(a) Scheme of a CI: 1D cut of potential energy surfaces along a generic reaction coordinate, $q$. Following photoexcitation from the ground state (GS), the wave packet reaches the CI from the Frank-Condon (FC) region. (b) Loop diagram \cite{loopDiagr} of the off-resonant stimulated Raman signal. The grey area represents the preparation of the system into the excited state by means of a pump pulse, temporally well separated from the detection process. After photoexcitation, the system propagates freely for a delay time $T$ before being probed by the hybrid-shaped pulse. (c) Schematics of the pump ($\varepsilon _p$) and hybrid probe pulse ($\varepsilon _1, \varepsilon ^{*} _0$) setup in TRUECARS.}
    \label{CI-scheme}
\end{figure}

Conical Intersections \cite{teller1937crossing, yarkony1998conical, worth2004beyond, domcke2011conical} (CIs) represent fast, radiationless decay channels in electronically excited molecules (see fig.\ \ref{CI-scheme}(a)).
Virtually present in every molecular system, CIs play a key role in charge transfer processes \cite{balzani2001electron}, reaction mechanisms \cite{zimmerman1966molecular} and in the vast majority of photochemical, photophysical and photobiological reactions \cite{klessinger1995excited, chergui2015ultrafast, robb2000computational,duan2016quantum, domcke2004conical, domcke2012role}, such as the \textit{cis}/\textit{trans} isomerization of retinal \cite{rinaldi2014comparison, polli2010conical}.
At such intersections the Born-Oppenheimer approximation breaks down causing complex dynamics
of the coupled vibronic states, which can be observed spectroscopically \cite{cederbaum1977strong, cederbaum1981multimode, young2018roadmap}.
As the photoexcited wave packet comes closer to the conical intersection, the energy separation between the potential energy surfaces (PESs) rapidly decreases. Thus, detection requires an unusual combination of temporal and spectral resolution that is not available via conventional femtosecond optical and infrared (IR) experiments \cite{polli2010conical, horio2009probing, Oliver10061, mcfarland2014ultrafast}. However, pulses in the extreme ultraviolet (XUV) to the short X-ray spectral region possess the required combination to directly detect the passage through a CI \cite{galbraith2017few, sun2020multi, yang2018imaging, wolf2017probing, chang2020revealing, zinchenko2021sub, nam2021conical, neville2018ultrafast, rohringer2019x}.
Single attosecond pulses (SAPs) have been extensively used in pump-probe experiments. Although the availability of such pulses has seen a significant increase thanks to high-harmonic generation (HHG) \cite{ferray1988multiple, varju2009physics} and free electron lasers (FELs) \cite{pellegrini2016physics} sources, the generation of SAPs still requires a complex setup.
The HHG process in gases emits a sequence of short bursts of radiation, which are coherently driven by the generation laser, where emission events occur during each laser half cycle.
Each of these short bursts is in the attosecond regime and their interference leads to the observation of odd harmonics.
Such attosecond pulse trains (APTs) \cite{farkas1992proposal, harris1993atomic}, unlike isolated pulses \cite{hentschel2001attosecond, paul2001observation}, are directly available in a HHG setup, which is a table-top source now widely found in many laboratories. Recent theoretical developments have showed the capability of APTs to probe the
electronic coherence created by the CI in the context of time-resolved photoelectron spectroscopy \cite{jadoun2021time} and time-resolved electron-momentum imaging \cite{Marciniak2019}.

In this paper, we demonstrate theoretically how APTs perform as probes in the framework of the transient redistribution of ultrafast electronic coherences in attosecond Raman signals (TRUECARS) technique \cite{TRUECARS} to probe electronic coherences generated by a wave packet passing through a CI. We use two different model systems that represent molecules of different symmetry and use quantum dynamics simulations to study the difference in the spectrum.

\section{Spectroscopic Signals and Models}
\subsection{The TRUECARS signal}\label{TRUECARS_section}
The TRUECARS technique uses an off-resonant stimulated X-ray Raman process (see fig.\ \ref{CI-scheme}(b) and (c)) which is sensitive to coherences rather than populations.
In the X-ray Raman scheme, core-hole states are involved as intermediates rather than common valence excited states.
As shown in fig.\ \ref{CI-scheme}(b), the two pulses making up the hybrid pulse scheme, namely $\varepsilon _1$ and $\varepsilon _0$, drive the Raman process, which is in turn
detected by a heterodyne detection scheme, where a local oscillator is used.
The frequency and time resolved signal reads in atomic units as follows:
\begin{align}\label{truecars_eq}
    S(\omega _R, T) = 2\Im \Bigg\{ \int \dd{t} \, & e^{i\omega _R (t-T)} \varepsilon ^{*} _0 (\omega _R) \nonumber \\
     \times & \varepsilon _1 (t-T)\langle \hat{\alpha}(t) \rangle \Bigg\}
\end{align}
where $T$ is the time delay between the probe field and the preparation pulse (see fig.\ \ref{CI-scheme}(c)), $\langle \hat{\alpha}(t) \rangle$ is the time dependent expectation value of the X-ray transition polarizability and $\omega _R$ is the Raman shift.
For the details of the signal, see Ref. \cite{TRUECARS}. The dependence on the dynamics of the system enters the TRUECARS signal through $\langle \hat{\alpha} \rangle$.
Additional details on $\hat{\alpha}$ are given in subsection C.

The TRUECARS spectrum is characterized by an oscillating pattern of gain and loss features in the Stokes and anti-Stokes regime, that is only visible when a vibrational or electronic coherence is present.
Assuming that both components of the probing field have the same carrier frequency, $\omega _X$, the spectrum is going to be centered at a Raman shift of $\omega_R = 0$.

\subsection{Modelling of the Pulse Trains}\label{Modelling_APT}
To illustrate how the pulse trains were built, we start from their definition in the time domain:
\begin{equation}\label{APT}
    E_{APT}(t) = G(t) P(t)
\end{equation}
where $G(t)$ is a Gaussian envelope of $\sigma _{env}$ width
\begin{align}
G(t) = e^{-t^2 /2\sigma _{env} ^2}
\end{align}
and $P(t)$ is an infinite train of pulses. The electric field of the single pulses inside the train is defined as
\begin{align}
E_{SAP}(t) = e^{-\left(t-\dfrac{\tau}{2}\right)^2 /2\sigma _{SAP} ^2} \cos\Big[{\omega _X \left(t-\dfrac{\tau}{2}\right)}\Big]
\end{align}
with $\omega _X$ being the center frequency and $\tau = 2\pi/\omega _{IR}$ the period of the IR field. The expression for the electric field then reads:
\begin{align}
& E_{APT}(t)  =   e^{-t^2 /2\sigma _{env} ^2} \nonumber \\
&\times \sum _n  e^{-\left(t-n\dfrac{\tau}{2}\right)^2 /2\sigma _{SAP} ^2}
             \cos \Big[{\omega _X \left(t-n\dfrac{\tau}{2}\right) + n\pi}\Big]\,. \label{APT_complt}
\end{align}
The pulse train employed in the calculations was built according to eq.\ \eqref{APT_complt} by substitution of $t$ with $t-T$ and with the following parameters: $\sigma _{env} = 2.5$\,fs and $\sigma _{SAP} = 0.15 $\,fs. For the purpose of simulating TRUECARS spectra, the center frequency $\omega _X$ can assume any arbitrary value.

\begin{figure}
    \includegraphics[width=\linewidth]{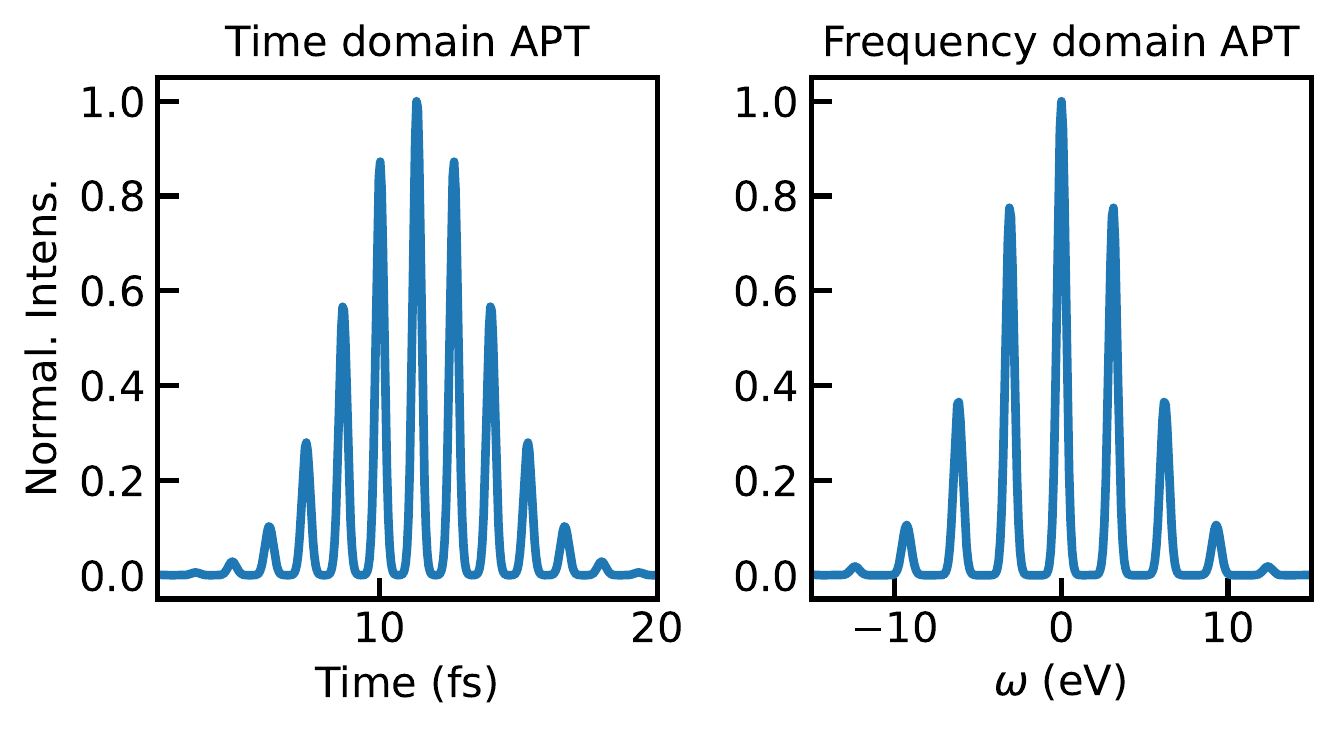}
    \caption{Snapshot of the modeled APT shape at $\omega _{IR} = 1.55$\,eV. On the left: envelope of the pulse train in time domain. On the right: pulse train in frequency domain, shifted with respect to a selected central harmonic.}
    \label{APT-train1-155}
\end{figure}

To ease nomenclature, henceforth, single pulses in the femtosecond or attosecond regime will be broadly referred to as "Gaussian pulses". A snapshot of the train pulse at $\omega _{IR} = 1.55$\,eV is displayed in fig.\ \ref{APT-train1-155}. Snapshots at different frequencies of the IR laser are available in the SI.

\subsection{Models and Symmetry}\label{Models}
We use group theory to identify Raman active transitions and the vanishing integral rule
to predict whether the polarizability matrix elements such as $\br{\Psi\dprime}  \hat{\alpha} \ke{\Psi\pr} $ will be zero.
In particular, they will vanish if the product of the irreducible representations of the two relevant states and the operator does not contain the totally symmetric
representation, that is
\begin{equation}\label{int_vanish_rule}
 \Gamma (\Psi \dprime) \otimes \Gamma (\hat{O}) \otimes \Gamma (\Psi \pr) \not\supset \Gamma ^{(s)} \, .
\end{equation}

The two systems studied in this work belong to the $C _i$ and to the $C_s$ point groups, respectively.
According to their character tables, the polarizability tensor elements $\alpha _{xx}$, $\alpha _{yy}$ and $\alpha _{xy}$ all transform as the totally symmetric irreducible representation, i.e. $A _g$ in $C _i$ and $A'$ in $C _s$. This was taken into account while modelling the X-ray polarizability operator, $\hat{\alpha}$, that was here approximated to a 2x2 matrix in the electronic subspace:
\begin{equation} \label{matrix-a}
        \hat{\alpha} = \begin{pmatrix}
\alpha _{11} & \alpha _{12}\\
\alpha _{21}& \alpha _{22}
\end{pmatrix} \,.
\end{equation}
The full electronic polarizability matrix must be taken into account in order to be basis-independent (adiabatic vs. diabatic states) and the diagonal matrix elements cannot be neglected when transforming between representations.
The diabatic wave function, $\Psi (\vec q, t)$, is expressed in terms of the electronic states. For a system with two electronic states, the wave function reads:
\begin{equation}
	\Psi(\vec q, t) = \begin{pmatrix} \phi _1 (\vec q, t)  \\ \phi _2 (\vec q, t) \end{pmatrix} \,.
\end{equation}
Adiabatic and diabatic states are related by an unitary transformation, where the advantage of the diabatic basis is the absence of derivative couplings \cite{kowalewski2017simulating}.
Expanding the expectation value, $ \langle \hat{\alpha} \rangle $, yields:
\begin{align}
    \langle \hat{\alpha} \rangle = &  \langle \Psi \vert  \hat{\alpha} \vert \Psi  \rangle \nonumber \\
     =& \langle \phi _1 \vert \alpha _{11}\vert \phi _1  \rangle + \langle \phi _2 \vert \alpha _{22}\vert \phi _2 \rangle
     + 2\Re\langle \phi _1 \vert \alpha _{12}\vert\phi _2 \rangle  \label{exp_val_polar_terms}
\end{align}
According to eq.\ \eqref{int_vanish_rule}, for a Raman active transition two conditions need to be fulfilled here:
i) the diagonal and off-diagonal polarizability matrix elements must all transform as the totally symmetric irreducible representation of their point group ($A _g$ in $C _i$ and $A'$ in $C _s$, respectively); ii) the electronic states $\phi _1$ and $\phi _2$ must have the same symmetry label, or else the integral $2\Re\langle \phi _1 \vert \alpha _{12}\vert\phi _2 \rangle $ will vanish.
The diagonal, ($\langle \phi _1 \vert \alpha _{11}\vert \phi _1  \rangle + \langle \phi _2 \vert \alpha _{22}\vert \phi _2 \rangle$), and off-diagonal, ($2\Re\langle \phi _1 \vert \alpha _{12}\vert\phi _2 \rangle $), terms refer to vibrational and electronic contributions, respectively.
Each vibrational normal mode has an associated irreducible representation, too. Therefore, the diagonal contribution will be suppressed if the vibrational modes transform as the wrong irreducible representation.
In practice, we can ensure the transition will be Raman active if $\phi _1$, $\phi _2$ and $\alpha _{ij}$ all fall into $A _g$ for $C _i$ and into $A\pr$ for $C _s$.

The concept of point groups is based on the approximation of a rigid molecule, that is considering the molecule as a rigid skeleton of nuclei. However, when large amplitude motions are considered where the symmetry of the system is not conserved, the concept is inadequate. Photoisomerization processes are well known examples of large amplitude nuclear motion in molecules.
To generate symmetry adapted polarizability element functions in the diabatic basis, we now introduce the complete nuclear permutation inversion (CNPI) group \cite{CNPI}. A CNPI group consists of all permutations of identical nuclei, the inversion of all nuclear and electronic coordinates, ($E^*$), as well as their products. The inversion $E^*$ differs from the inversion operation $i$ in point groups. The former is an operation which results in a sign change of all nuclear and electronic coordinates in the space-fixed coordinate system.
 A similar application to electronic states and transition dipole moments can be found in Ref. \cite{obaid2015molecular}. Following these sections, details of the modeling of the two systems as well as the polarizability matrix are shown.

\subsubsection{\texorpdfstring{$C_i$}{Ci} Symmetry Model}\label{BS_sect}
\begin{figure}
    \includegraphics[width=\linewidth]{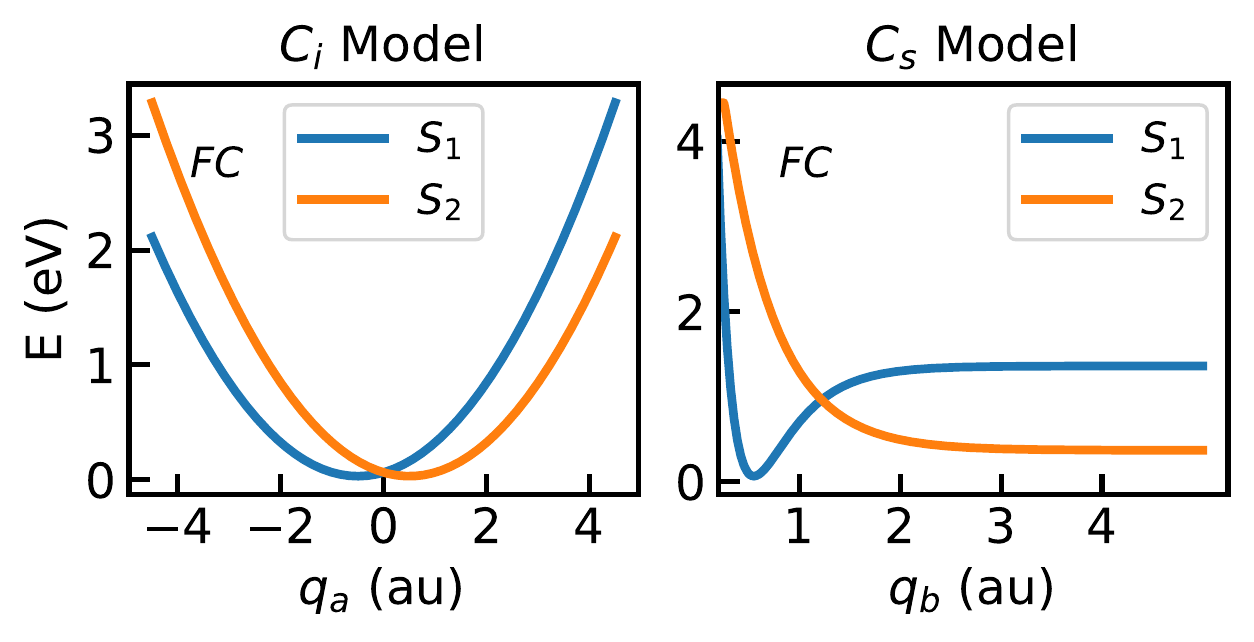}
    \caption{Potential energy surfaces of the investigated model systems. On the left: 1D cut of the diabatic potential energy surface along $q _a $, in the $C_i$ system.  On the right: 1D cut of the diabatic potential energy surface along $q _b$, in the $C_s$  system.
    The wave packet  has been previously excited from the ground state to the Franck-Condon (FC) region on $S _2$ by means of a pump pulse.}
    \label{1D-PES}
\end{figure}

This model consists of two harmonic potential wells shifted with respect to each other (see left panel of fig.\ \ref{1D-PES}).
It represents two excited electronic states with a conical intersection. An example of such a molecule could be benzene (C$_6$H$_6$) and its photochemistry \cite{palmer1993mc}  or acetylene (C$_2$H$_2$) \cite{KOPPEL2008319}.
The diagonal and off-diagonal elements of the polarizability have been shaped by a function of $q_1$ and $q_2$, which is symmetric with respect to the two normal modes. In the framework of CNPI this translates as
\begin{equation*}
    E^* \alpha _{ij} (q_1, q_2) = E^* \alpha _{ij}  (-q_1, -q_2) = E^* \alpha _{ij}  (q_1, q_2) \,.
\end{equation*}
This function behaves non-linearly with respect to the nuclear coordinates in the proximity of the CI. In this specific case, both normal modes $q_1$ and $q_2$ transform as the totally symmetric irreducible representation $A_g$, thus they are all Raman active.
Plots of the symmetry adapted  functions of the polarizability matrix elements as well as the single contributions to $\ev{\hat{\alpha}}$ are available in the SI.

\subsubsection{\texorpdfstring{$C _s$}{Cs} Symmetry Model}\label{UM_sect}
The $C_s$ model consists of a Morse-like potential well and a repulsive potential with asymptotic behavior at large values of $q$, with a conical intersection. A 1D cut of the system is shown in the right panel of fig.\ \ref{1D-PES}. Hydroxylamine (NH$_2$OH), which has $C _s$ symmetry, is a  well-known example for the study of the effects of conical intersections in photodissociation \cite{RevModPhys.68.985}.
Similarly to the $C_i$ symmetry, the diagonal and off-diagonal elements of the polarizability have been here shaped by a function of $q_1$ and $q_2$. This time, the function is symmetric with respect to $q_2$ and non-symmetric to $q_1$.
A fundamental transition is Raman active if a normal mode forms a basis for one or more components of the polarizability. Here, the normal mode $q_1$ transforms as the irreducible representation $A\dprime$, while $q_2$ as $A\pr$. Thus, $q_2$ is Raman active. Further details, including contour plots of the aforementioned functions, are available in the SI.

\section{Computational Details}\label{Comput_Det_sect}\
The time evolution was simulated by solving the time-dependent (non-relativistic) Schr\"odinger equation
numerically with the Fourier method \cite{tannor}, where the wave function is represented on an equally spaced grid of sampling points in coordinate space, using the in-house software QDng. In the diabatic picture, the two-level Hamiltonian reads as:
\begin{equation} \label{gs-Hamiltonian}
    \hat{H} = \begin{pmatrix}
            \hat{V}_1 + \hat{T} & \hat{V}_{12} \\
            \hat{V}_{12} & \hat{V}_2 + \hat{T}
             \end{pmatrix} \, ,
\end{equation}
where $\hat{T}$ is the kinetic energy operator
\begin{equation}
\hat{T} = -\frac{\hbar ^2}{2\mu}\sum _i \nabla _{q_i} ^2
\end{equation}
while $\hat{V}_1$ and $\hat{V}_2$ are the potential energy operators, respectively for $S_1$ and $S_2$, and $\hat{V}_{12}$ is the diabatic coupling operator.
The Arnoldi scheme \cite{arnoldi1951principle, friesner1989method, saad1980variations} was employed as the propagator for all calculations. The reduced masses, $\mu$, the time steps, $\Delta t$, and the number of grid points employed have been all summarized in table \ref{param_calc}.  A $\Delta t$ of 4 is equal to 96.75 as, while a reduced mass of 18000 is approximately 10 amu. For the initial state, assumed to be the result of a short-fs excitation pulse, the nuclear wave packet was approximated by a Gaussian envelope on $S _2$. Lastly, a perfectly matched layer \cite{nissen10jcp} was employed to absorb the wave packets at the boundary and to account for the dissociative behaviour.
\begin{table}[htp]
\begin{tabular}{lcccc}
          &  & $\mu$ (au) & $\Delta t$ (au) & Grid p. \\ [3pt]\hline
$C _i$ &  &  18000   &     4   &  256x256  \\
$C _s$ &  &  30000  &       2   & 300x300 \\ \hline
\end{tabular}
\caption{Reduced masses, time steps and number of grid points used in the simulations.}
\label{param_calc}
\end{table}
The propagated wave packets were used for the evaluation of the expectation value of the polarizability, $\langle \hat{\alpha} \rangle$. Finally, the TRUECARS spectra were calculated with eq.\  \eqref{truecars_eq} at different values of the pump-probe delay, $T$.

\section{Results}\label{Results}

\begin{figure}
\includegraphics[width=\linewidth]{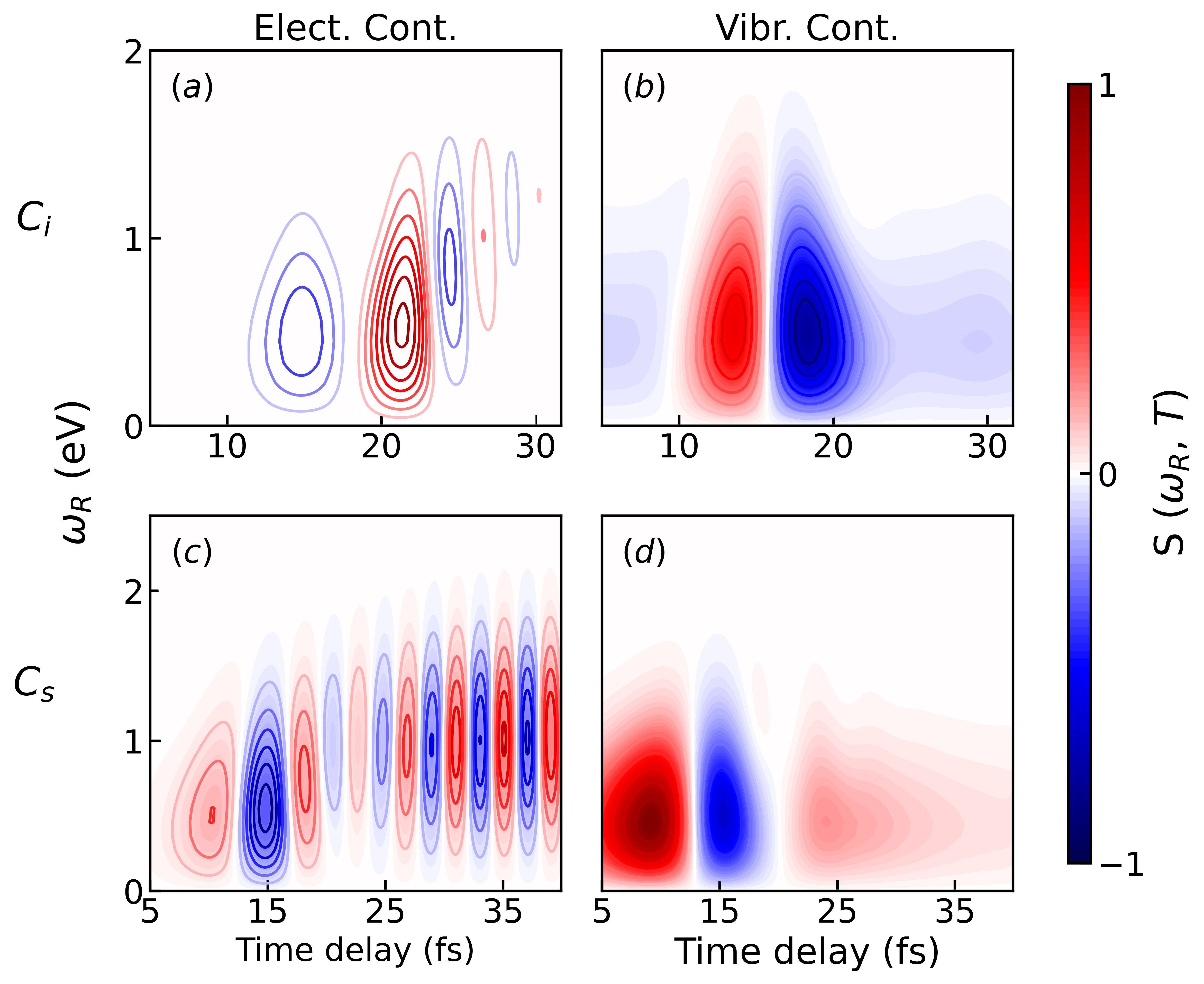}
    \caption{Comparison of electronic and vibrational contributions to
    the TRUECARS signals for Gaussian pulses.
    (a): electronic contribution in the $C_i$ model; (b): vibrational contribution in the $C_i$ model; (c): electronic contribution in the $C_s$ model; (d): vibrational contribution in the $C_s$ model. Spectra in the same row share the same Raman shift axis. The signals have been normalized with respect to the maximum value of (d).}
    \label{Ci-Cs-2contribut}
\end{figure}

For each system we have simulated two spectra with the conventional TRUECARS probe scheme, containing purely electronic or vibrational signals, as displayed in fig.\ \ref{Ci-Cs-2contribut}. This was achieved by simulating the vibrational contribution
($\br{\phi_1} \alpha _{11}\ke{\phi_1}
+ \br{\phi_2}\alpha_{22}\ke{\phi_2}$)
and the electronic contribution ($2\Re\br{\phi_1} \alpha_{12}\ke{\phi_2}$) separately.
This distinction will help us break down the interesting characteristics of the spectra and highlight the differences and similarities between the probe schemes.
It should be emphasized that the distinction of vibrational and electronic degrees of freedom near a conical intersection is fictitious, as these degrees of freedom are mixed.

Signals in fig.\ \ref{Ci-Cs-2contribut} have been normalized with respect to the maximum value of the vibrational contribution in the $C_s$ model (panel (d)) with the following ratios: (a) $2\cdot 10^{-4}$, (b) 0.78 and (c) 0.33. By comparing the overall intensities between the two models, we notice that the electronic contribution in the $C_s$ is higher than the $C_i$ system.
The vibrational signals are, in both model systems, contained within the $ 0\leq \omega _R \leq 1.5$ eV range and immediately visible in the spectrum due to vibrational coherences ($\langle \phi _2 \vert \alpha _{22}\vert \phi _2 \rangle$), whereas the electronic signals spread over a broader energy range ($ 0\leq \omega _R \leq 2$ eV) and only appear after the wave packet has reached the CI. Hence, if the energy resolution is not sufficient, the two will overlap with each other and the electronic component may be masked by the stronger vibrational signal. A major difference between the two components can be seen in the temporal oscillations, where the electronic contribution to the signal oscillates much faster than the vibrational one.

In contrast to the conventional hybrid probe, here pulse trains have been employed instead of single Gaussian pulses. The following three combination schemes have been investigated:
\begin{itemize}
    \item[$(I)$] An APT as $\varepsilon _1$ and a Gaussian pulse as $\varepsilon ^* _0$;
    \item[$(II)$] A Gaussian pulse as $\varepsilon _1$ and an APT as $\varepsilon ^{*} _0$;
    \item[$(III)$] Two identical APTs as both fields.
\end{itemize}
In the following, we use $\sigma _0$ and $\sigma _1$ to indicate the Gaussian width of $\varepsilon ^{*} _0$ and $\varepsilon _1$, respectively.
Due to the nature of the pulse trains, multiple signals are expected in the spectra simulated with schemes $(I)$ and $(III)$. The extra signals arise from the side peaks of the train and are expected to be symmetric to each other, but they are lower in intensity with respect to the central signal at $\omega _R = 0$. In scheme $(II)$, a Gaussian pulse in the femtosecond timescale is employed as $\varepsilon _1$ and the spectrum will mostly consist of one central signal. This is due to the limited spectral resolution of the femtosecond narrow band pulse. Spectra similar to fig.\ \ref{Ci-Cs-2contribut} but simulated with probe scheme $(III)$, are given in the SI.

Calculations were carried out for different values of $\omega _{IR}$ at 1.55, 0.99 and 0.83 \,eV (i.e., $\lambda _{IR} =$ 800, 1250, 1500 nm). The IR laser frequency is an important parameter in these calculations because it directly shapes the attosecond pulse train via the IR laser period, $\tau$. A smaller $\tau$ implies more peaks in the time domain and, equivalently, less peaks in the frequency domain. Similarly, the $\sigma _{env}$ parameter in eq.\ \eqref{APT_complt} can achieve the same effect.

\subsection{\texorpdfstring{$C_s$}{Cs} Symmetry Model}
\begin{figure}
    \includegraphics[width=\linewidth]{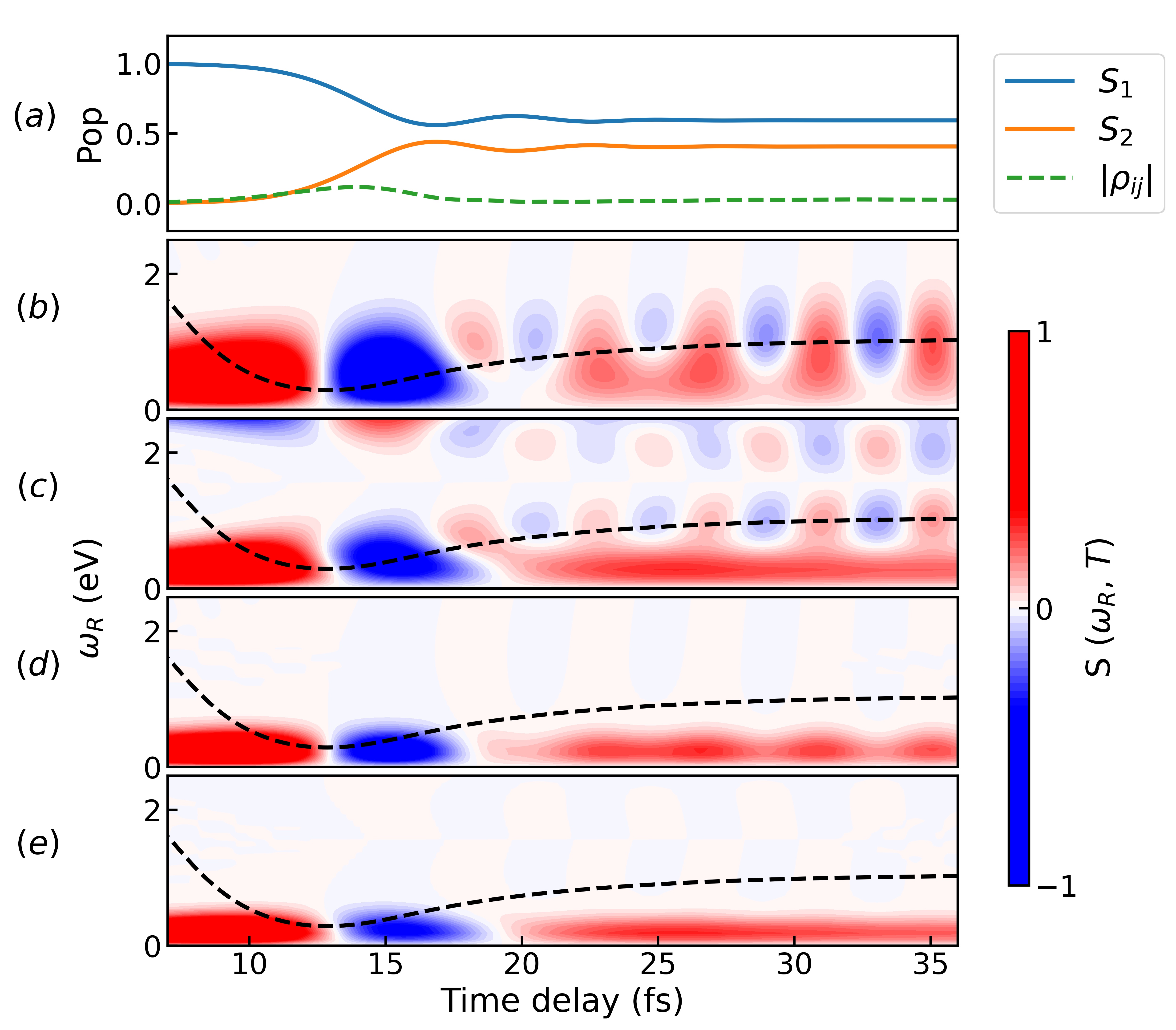}
    \caption{Comparison of pulse schemes for the $C _s$ model system.
    (a): Time evolution of the population and the coherence of the two excited states in the diabatic basis. The population transfer is around 45$\%$ and occurs after about 15 fs of the wave packet propagation. The diabatic coupling is responsible for the slight oscillation in the population between 15 and 20 fs; (b) TRUECARS spectrum generated by an isolated Gaussian hybrid femtosecond/attosecond probe-pulse sequence (pulse parameters $\sigma  _1 = 1.5$ and $\sigma _0 = 0.15$\,fs). (c) TRUECARS spectrum obtained via combination $(I)$ with Gaussian pulse parameter $\sigma _0 = 0.15$\,fs; (d) TRUECARS spectrum obtained via combination $(II)$  with Gaussian pulse parameter $\sigma  _1 = 1.5$\,fs; (e) TRUECARS spectrum obtained via combination of two identical APTs (scheme $(III)$). Each signal has been normalized with respect to its maximum value. The dashed black line represents the average time-dependent separation of the adiabatic potential energy surfaces.
    All spectra are simulated for $\omega _{IR} = 1.55$ eV. A snapshot of the pulse train can be found in fig.\ \ref{APT-train1-155}.}
    \label{Cs-Truecars}
\end{figure}

We begin our discussion of the results starting with the $C _s$ model. The time evolution of the population of the excited states is plotted in the top panel of fig.\ \ref{Cs-Truecars}. Following photoexcitation, the wave packet reaches the CI in $\approx$12 fs with an overall population transfer of $\approx$45\%. The electronic coherence reaches a maximum of 0.15 at 15\,fs, after which starts decaying. The simulated APT TRUECARS spectra are shown in fig.\ \ref{Cs-Truecars} (c), (d) and (e) compared to a standard single pulse TRUECARS (b). The dashed black line in the spectra indicates the expectation value of the energy separation, $\Delta E _{21} (t)$, between $S_1$ and $S_2$. For more details, see the S.I. of Ref. \cite{TRUECARS}. Because the spectrum is symmetric with respect to $\omega _R = 0$, only signals within $\omega _R \in [0,2.5]$ eV are shown. Extra signals appear above 2.5 eV with schemes $(I)$ and $(III)$, however, those only carry redundant information, as they are lower-intensity replicas of the central peak. Initially, the vibrational coherence is the only contribution and it is constrained within $ 0\leq \omega _R \leq 1.5$ eV, as shown in fig.\ \ref{Ci-Cs-2contribut}. Once the wave packet is in the proximity of the CI, the electronic coherence starts to build up and becomes visible in the $ 0\leq \omega _R \leq 2$ eV region of the spectrum. As the energy separation between the states increases again, the oscillating pattern of gain and loss features in the Stokes and anti-Stokes regime can be seen in the spectrum. The oscillation period directly mirrors the coherence period: as $\Delta E _{21} (t)$ grows, the oscillations speed up which causes the positions of the peaks in the Raman shift $\omega _R$ to spread apart. Due to the shape of the potential energy surfaces, the energy separation between $S _2$ and $S _1$ stays approximately constant after the CI. This can be seen from the dashed black lines in fig.\ \ref{Cs-Truecars} as well as the positions of the peaks in the Raman shift $\omega _R$.

Among the three APT probe combinations displayed in fig.\ \ref{Cs-Truecars} only (c) is able to capture the CI signature at $\omega _{IR} = 1.55$ eV. By comparing panels \ref{Cs-Truecars}(b) and (c), we note that scheme $(I)$ can achieve similar energy resolution to the Gaussian/Gaussian hybrid probe, while being characterized by the presence of additional signals in the spectrum. However, probe scheme $(I)$ still requires a single pulse in the attosecond time scale.

\begin{figure}
    \includegraphics[width=\linewidth]{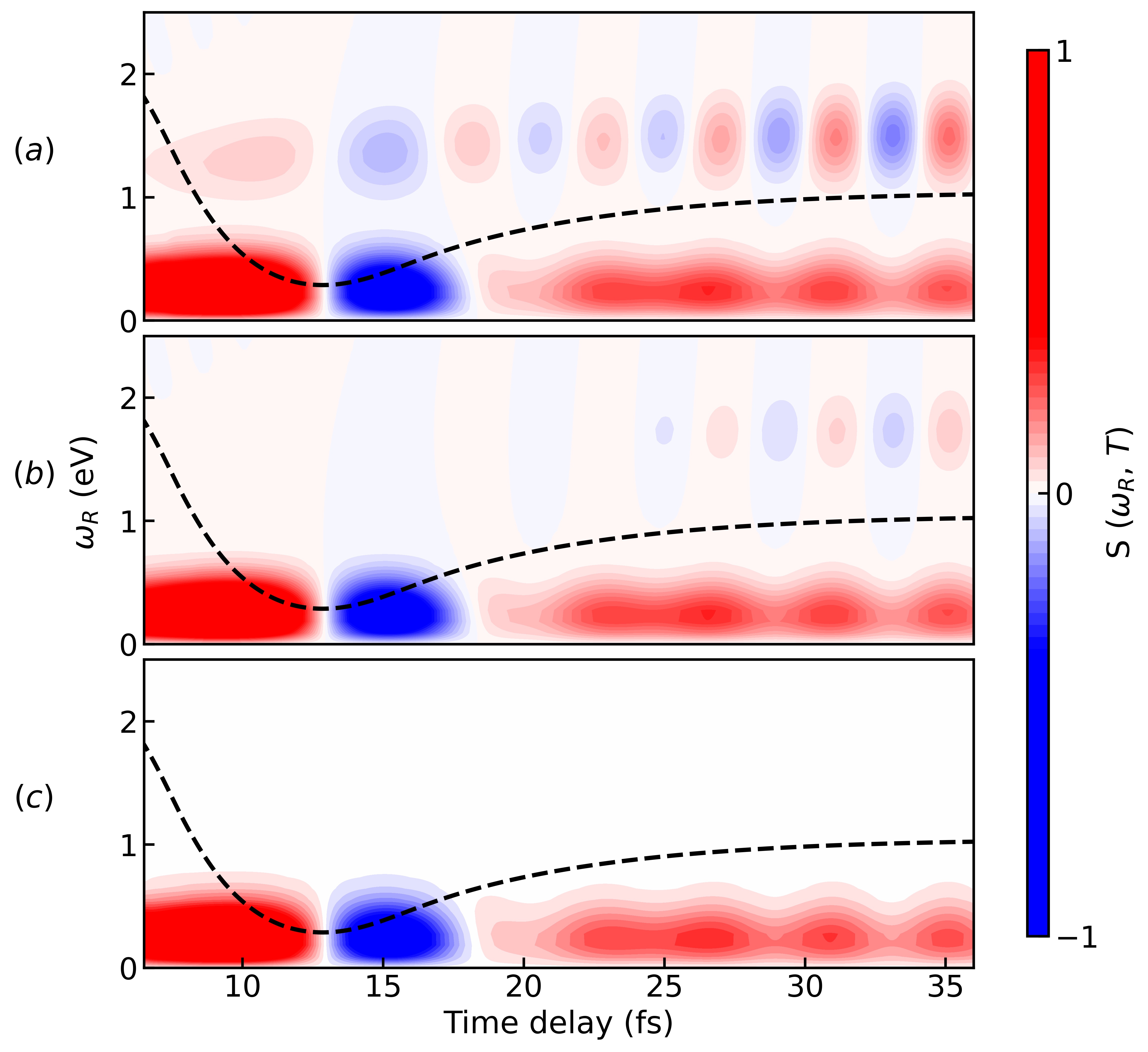}
    \caption{Simulated TRUECARS spectra for increasing values of the generating IR laser frequency with probe scheme $(II)$, for the $C_s$ system. (a): $\omega_{IR} = 0.83$\,eV;  (b): $\omega_{IR} = 0.99$\,eV; (c): $\omega_{IR} = 1.55$\,eV. The Gaussian pulse employed as $\varepsilon_1$ has a width of $\sigma_1 = 1.5$\,fs. The dashed black line represents the average time-dependent separation of the adiabatic potential energy surfaces.}
     \label{Cs-freqs-TRUECARS}
 \end{figure}

As the IR laser frequency decreases from 1.55 to 0.83 eV, the pulse train peaks get closer to each other in the spectral domain and the interesting electronic coherence fingerprint becomes visible in the spectra simulated with scheme $(II)$, as displayed in fig.\ \ref{Cs-freqs-TRUECARS}. This probe scheme does not require a single pulse in the attosecond regime. Moreover, it allows to achieve a better spectral resolution of the signals in the Raman shift than a standard TRUECARS (fig.\ \ref{Cs-Truecars}(b)). In fact, the vibrational and electronic contributions are now sufficiently separated from each other to allow for an unambiguous assignment. The latter is not captured by the central harmonic of the APT but by its first side peak. This hypothesis was supported by simulations carried out varying the $\sigma _{env}$ parameter, at the same $\omega _{IR}$ of fig.\ \ref{Cs-freqs-TRUECARS}. The width of the Gaussian envelope directly shapes the width of the harmonics in the HHG spectrum. By increasing and decreasing the $\sigma _{env}$ of the pulse train we noticed a corresponding increase and decrease in the separation between harmonics, and therefore between the vibrational and electronic contribution in the TRUECARS signal. The oscillation in fig.\ \ref{Cs-freqs-TRUECARS} appears to be slightly shifted in the Raman shift and does not follow the dashed black line anymore. Nevertheless, the oscillation period is the same and can still be used to obtain information on CI. According to the calculations, for this specific system an IR laser frequency of 0.83 eV appears to be the most suited to resolve the time-dependent energy gap between the two PESs, as shown in fig.\ \ref{Cs-freqs-TRUECARS}(a).

\begin{figure}
    \includegraphics[width=\linewidth]{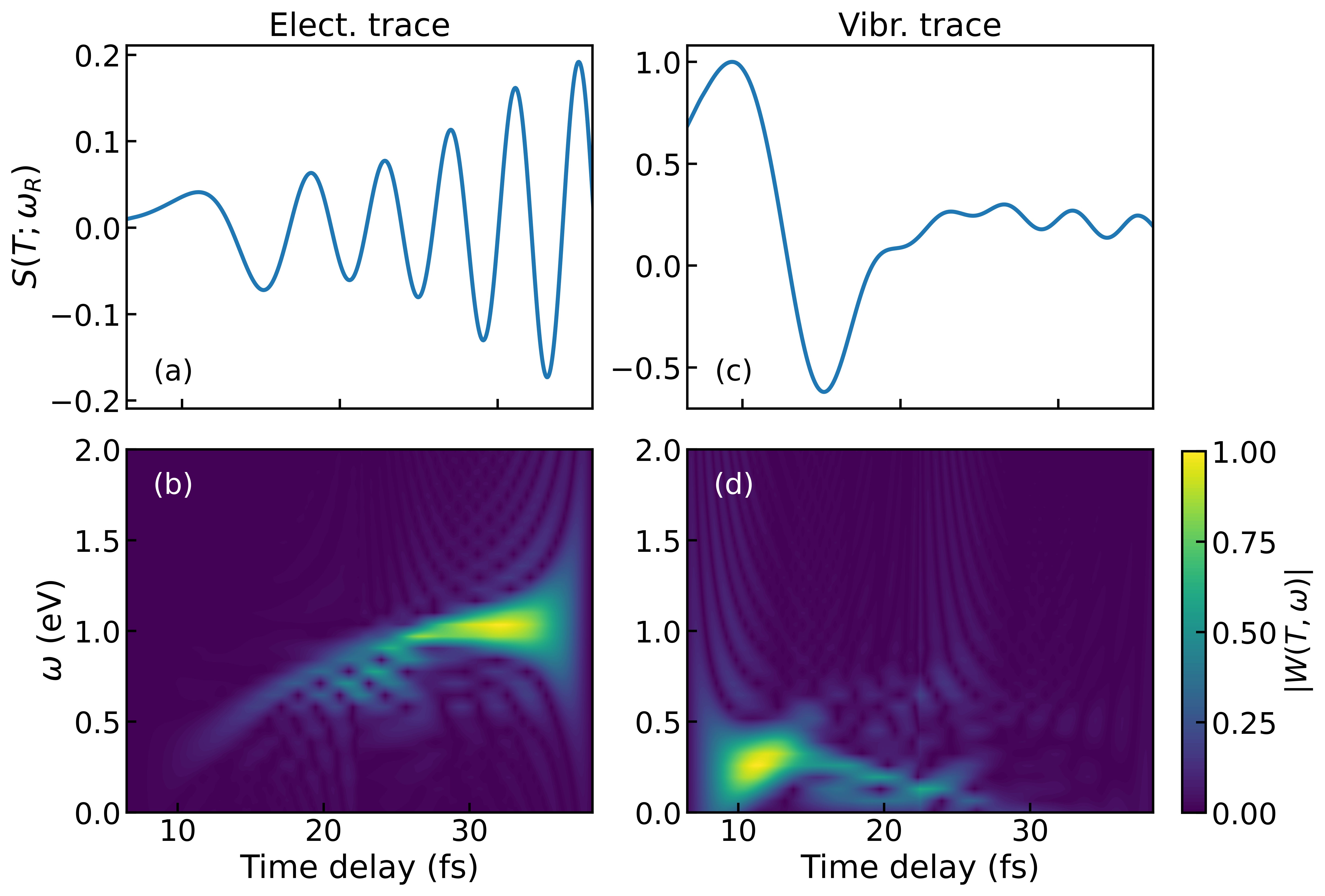}
    \caption{Comparison between Wigner distributions of selected traces of the TRUECARS signal. (a) and (c): Signal traces $S(T;\omega _R)$ at $\omega _R = 1.56$ and 0.27 eV, respectively. (b) and (d): Normalized Wigner spectrograms of (a) and (c). The frequency information obtained from the spectrogram is extracted from the temporal features of the signal.}
    \label{Wigner_spectr}
\end{figure}

We can extract additional insight from the oscillation period by analyzing the Wigner distribution \cite{keefer2020visualizing, cavaletto2021high, cavaletto2021unveiling} or temporal gating spectrogram of the analytic signal \cite{boashash1988note}, which display a time-frequency map. The Wigner distribution is defined as
\begin{equation} \label{Wigner}
    W(T,\omega) = \int _{-\infty} ^{\infty} \dd \tau \, S_a\left(T + \dfrac{\tau}{2}\right) S_a^*\left(T - \dfrac{\tau}{2}\right) \, e^{i \omega \tau} \;
\end{equation}
where $S_a(T)$ is the so-called analytic signal whose imaginary part is related to the original signal, $S(T)$, by Hilbert transformation:
\begin{equation}
    S_a(T)= S(T) + \dfrac{i}{\pi}\bigintssss _{-\infty} ^{\infty} \dd s \, \dfrac{S(T-s)}{s}
\end{equation}
Here, $S(T)\equiv S(T; \omega_R)$ is a temporal slice of the signal in  fig.\ \ref{Cs-freqs-TRUECARS}(a) at a selected Raman shift.
The Wigner distribution is a quadratic functional of the signal and so it will, in general, show interference between the negative and positive frequency components of the signal. However, when the analytic signal is used in the computation, no negative frequencies are present, hence no interference will survive in the spectrogram.
Figure\ \ref{Wigner_spectr}(b) and (d) show the modulus $\vert W(T,\omega) \vert$ for signal traces taken at $\omega_R = 1.56$ and 0.27\,eV, which are interpreted as electronic and vibrational contributions respectively.
Panels\ \ref{Wigner_spectr}(a) and (c) capture the different temporal oscillations of the vibrational and electronic components of
the TRUECARS signal. The electronic Wigner distribution spans a broader frequency window than the vibrational contribution. Such window represents the PESs splitting in proximity of the CI. More strikingly, the energy splitting of the main electronic feature is time-dependent and starts around 0.25\,eV at 10\,fs and converges to 1\,eV at
$\approx 25\,$fs. This is in good agreement with the indicated
splitting (black dashed line) in figs.\ \ref{Cs-Truecars} and\ \ref{Cs-freqs-TRUECARS}.

With probe scheme $(III)$ (fig.\ \ref{Cs-Truecars}(e)) the characteristic features caused by electronic coherence are concealed in the spectrum. This happens because the small oscillation, traceable to the electronic component of the TRUECARS signal, overlaps with the dominant vibrational contribution in the same region.

\subsection{\texorpdfstring{$C_i$}{Ci} Symmetry Model}
\begin{figure}
   \includegraphics[width=\linewidth]{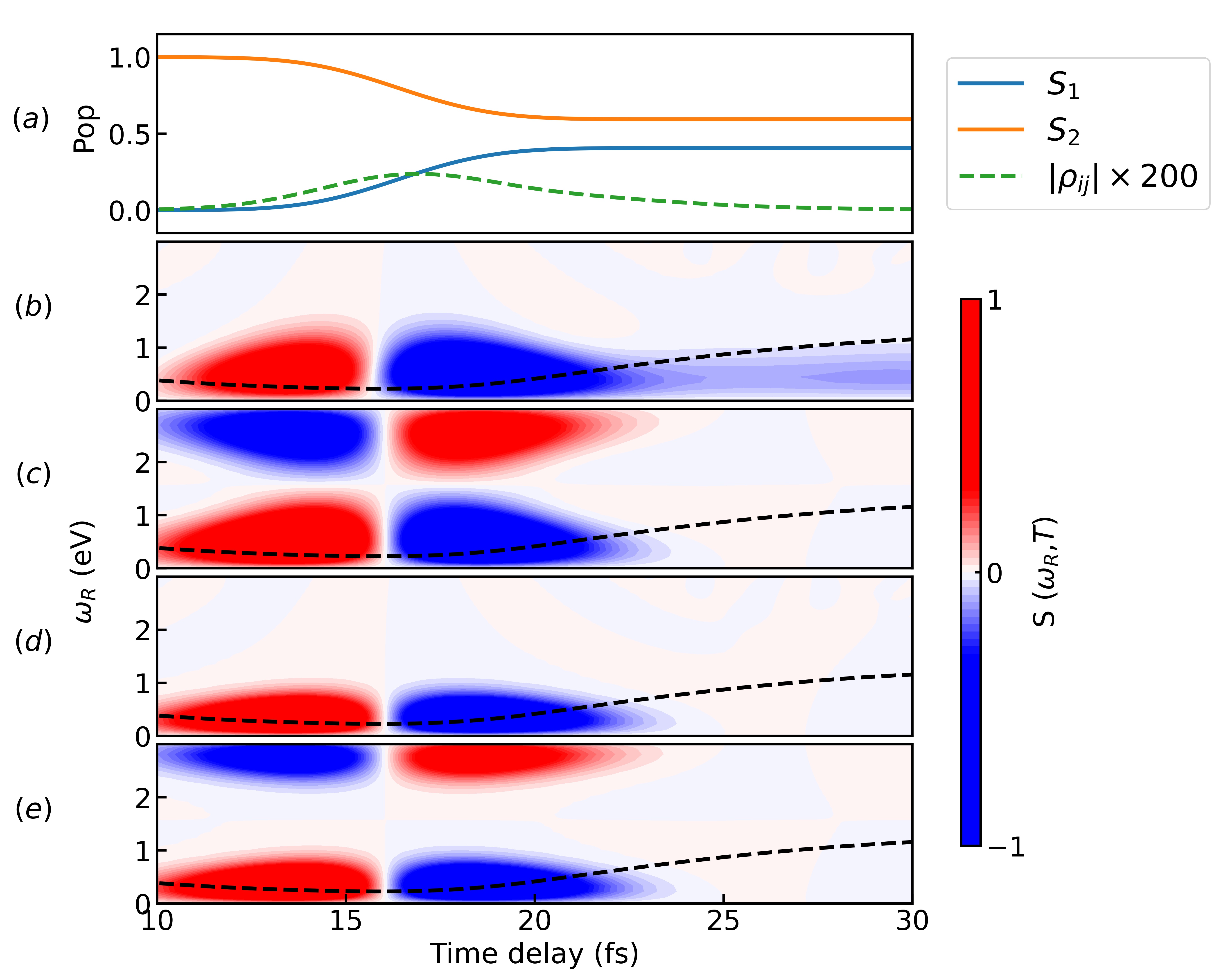}
    \caption{Comparison of pulse schemes for the $C _i$ model system.
      (a): Time evolution of the populations and coherence for states $S_1$ and $S_2$ in the diabatic basis. The coherence magnitude (in green) has been magnified by a factor of 200 for visual purposes. The population transfer ($\sim 45 \%$) occurs after about 15\,fs when the wave packet reaches the CI. (b): TRUECARS spectrum generated by an isolated Gaussian hybrid femtosecond/attosecond probe-pulse sequence (pulse parameters $\sigma  _1 = 1.5$ and $\sigma _0 = 0.15$\,fs). (c): TRUECARS spectrum obtained via combination $(I)$ with Gaussian pulse parameter $\sigma _0 = 0.15$\,fs; (d): TRUECARS spectrum obtained via combination $(II)$  with Gaussian pulse parameter $\sigma  _1 = 1.5$\,fs; (e): TRUECARS spectrum obtained via combination $(III)$. The dashed black line represents the average time-dependent separation of the adiabatic potential energy surfaces.
     All are spectra are calculated at $\omega _{IR} = 1.55$\,eV. Each signal has been normalized with respect to its maximum value. A snapshot of the pulse train is displayed in fig.\ \ref{APT-train1-155}.}
    \label{Ci-4plots_spectra}
\end{figure}

 The time evolution of the population of the two excited states, $S _2$ and $S _1$, as well as the electronic coherence magnitude, is displayed in fig.\ \ref{Ci-4plots_spectra}(a) for the $C_i$ model system.
The wave packet takes about 15\,fs to reach the conical intersection, resulting in an overall population transfer of $\sim$45\%. The electronic coherence has a maximum of 0.0015 around 15\,fs, after which starts decaying because of the increasing energy splitting between the surfaces.

 Following the photoexcitation on $S_2$, as the original wave packet approaches the CI, the wave packet transferred on $S_1$ will inherit an odd symmetry from the diabatic couplings. This generates a very weak electronic coherence ($10^{-3}$ order of magnitude) because the integral $ 2\Re\langle \phi _1 \vert \phi _2 \rangle$ gets very small. Hence, the vibronic coherence, which TRUECARS is sensible to via the physical observable $\langle \hat{\alpha} \rangle$, has now a dominant vibrational component that totally conceal the interesting and characteristic features of the CI, such as the time-dependent energy splitting. This is why we are not able to observe them with the TRUECARS technique for this model system, not even with the standard single Gaussian pulse probe scheme of fig.\ \ref{Ci-4plots_spectra}(b). Decreasing the IR frequency from $\omega _{IR} = 1.55$ to $0.83$ eV does not produce any significant change.
The reason why this does not occur in the $C_s$ model is due to the small shift of $S_2$ in $q_2$, breaking the symmetry and making the integral larger.

\section{Conclusions}
In this paper, we tested the suitability of attosecond pulse trains as probes for detecting electronic coherence generated at a conical intersection with the TRUECARS technique.
Model systems of different symmetry and two nuclear degrees of freedom were used. The polarizability matrices were modelled to obtain Raman active vibrational modes. The inclusion of the diagonal polarizability matrix elements includes vibrational coherences, which are inevitably a part of the signal. The full vibronic polarizability matrix must be taken into account in order to be basis-independent and the diagonal matrix elements cannot be neglected when transforming between adiabatic and diabatic representations. Although this leads to a more complex signal, we could show that it is still possible to distinguish the fast oscillating electronic feature from the vibrational contribution by analyzing the temporal gating spectrogram.

To gain additional insight into the TRUECARS signals, spectra originated from purely electronic or purely vibrational contributions were simulated by evaluating the off-diagonal and diagonal contribution separately in the time-dependent expectation value of the polarizability operator. Due to symmetry and Raman selection rules, the electronic coherence appears to vanish in the $C_i$ model system and concealed by the much stronger vibrational contributions.

We have explored three different schemes for the probe pulse and discussed their features in comparison to the conventional TRUECARS scheme.
We found that, among the schemes reviewed, the combination of a Gaussian pulse as a narrowband pulse and an attosecond pulse train as the broad band pulse proved to be the most suitable for our purposes, offering two main advantages: first, a more clear separation between the electronic and vibrational components of the TRUECARS signal can be achieved by fine-tuning the IR laser frequency. The convolution of harmonics leads to a shift of the peaks in the spectral domain, but the energy separation can still be read off the oscillation period in the time domain. This was corroborated by the analysis of Wigner spectrograms, calculated for selected temporal traces of the signal. Second, single Gaussian pulses in the attosecond timescale are no longer required. Furthermore, this is the logical choice for the hybrid probe used in TRUECARS, since the Gaussian pulse and the APT represent a narrowband and a broad band pulse, respectively. The calculations showed the best results at $\omega _{IR} = 0.83$\,eV.

The probe scheme with an APT and a Gaussian pulse, employed as narrowband pulse and broad band pulse respectively, also proved to be suitable to capture the CI fingerprints. Whereas the traditional TRUECARS spectrum is composed of one main signal centered at $\omega_R =0$, multiple bands are now visible. Such bands are replicas of the main central signal with scaled intensities, and are visible along the harmonic comb where the side peaks of the pulse train appear. When compared, the two spectra contain similar information about the detected vibronic coherence. Nevertheless, such combination for the probe still requires a single attosecond pulse acting as a broadband pulse.

The use of two identical pulse trains did not reveal most features related to the electronic coherence in the simulated spectrum.
Due to the insufficient resolution, the electronic component overlaps with the dominant vibrational signal, resulting in a concealment of the  electronic contribution.

\section*{Supplementary Material}
The supplementary material provides:  \textit{i}) snapshots of the modeled attosecond pulse train shape at selected frequencies of the generating IR laser; \textit{ii}) details about modeling of the material properties as well as the  potential energy surfaces  and the diabatic couplings used for the $C _i$ and $C_s$ systems; \textit{iii}) TRUECARS spectra displaying the entire frequency comb of the attosecond pulse train.

\begin{acknowledgements}
This project has received funding from the European Union's Horizon 2020 research and innovation program under the Marie Sk\l{}odowska-Curie grant agreement No. 860553 and the Swedish Research Council (Grant No. VR 2018-05346).
\end{acknowledgements}

\section*{Data Availability}
The data that support the findings of this study are available from the corresponding author upon reasonable request.

\bibliography{biblio.bib}

\bibliographystyle{unsrt}

\end{document}